\documentclass[final,3p,times,twocolumn,authoryear]{elsarticle}

\usepackage{amssymb}

\def\simlt{\mathrel{\rlap{\lower 3pt\hbox{$\sim$}}\raise 2.0pt\hbox{$<$}}}
\def\simgt{\mathrel{\rlap{\lower 3pt\hbox{$\sim$}} \raise 2.0pt\hbox{$>$}}}
\def\lsim{\mathrel{\rlap{\lower 3pt\hbox{$\sim$}}\raise 2.0pt\hbox{$<$}}}
\def\gsim{\mathrel{\rlap{\lower 3pt\hbox{$\sim$}} \raise  2.0pt\hbox{$>$}}}

\def\Msun{{\rm M}_{\odot}}
\def\Zsun{{\rm Z}_{\odot}}

\def\aap{A\&A}
\def\apj{ApJ}

\def\apjl{ApJ}
\def\mnras{MNRAS}
\def\araa{ARA\&A}

\def\nat{Nature}

\def\apjs{ApJS}

\def\ssr{ssr}
\def\pasj{PASJ}

\journal{Journal of High Energy Astrophysics}

\begin{document}

\begin{frontmatter}



\title{High redshift Gamma-ray Bursts}


\author{Ruben Salvaterra}

\address{INAF/IASF-Milan, via E. Bassini 15, I-20133 Milano, Italy}
\address{email: ruben@lambrate.inaf.it}

\begin{abstract}
Ten years of operations of the {\it Swift} satellite have allow us to
collect a small sample of long Gamma-Ray Bursts (GRBs) at redshift
larger than six. I will review here the present status of this
research field and discuss the possible use of GRBs as a fundamental new
tool to explore the early Universe, complementary to quasar and galaxy surveys. 
\end{abstract}


\end{frontmatter}


\section{Introduction}\label{sec:intro}

In the standard $\Lambda$CDM scenario, the first stars, the so-called
Population III (PopIII) stars, are predicted to
form in dark matter minihalos of typical mass $\sim 10^6\;\Msun$ at
$z\sim 20-30$ out of a gas of pristine composition \citep[see][for recent reviews]{Ciardi2005,BrommYoshida2011}. Thanks to their peculiar chemical
composition, they are expected to be more massive than the subsequent
stellar populations with typical mass of $\sim 40\;\Msun$ \citep{Hosokawa2011},
possibly extending up to hundred solar masses \citep{Hirano2014}. The formation of
these stars mark the fundamental transition from a simple, very
homogeneous Universe to the complex and structured one  we see already
in place a billion years after the Big Bang. During this period of
time two fundamental transitions are
expected to occur: (i) a change in the SFR mode, i.e. from massive 
PopIII to solar-size PopII/I stars and (ii) the cosmic
reionization, i.e. the change  of the inter-galactic
medium (IGM) from a neutral
to a fully ionized state. There is a general consensus that the first
transition is driven by the so-called chemical feedback
\citep{Schneider2002}, i.e. the enrichment of
star forming clouds by the first supernova explosions above a critical threshold of $Z_{\rm crit}=10^{5\pm
  1}\;\Zsun$ \citep{Schneider2003}. As the chemical feedback is
essentially a local effect, on cosmological scale the two populations
should be coeval for a long period of time \citep{Schneider2006,Tornatore2007,Maio2010}. Understanding how and when the change in the SFR mode happened
is one of the main goals for current studies of galaxy formation in
the early Universe and the detection of PopIII stars one of the main
challenges of the next generation of space and ground facilities
\citep[see][for a review]{BrommYoshida2011}. The cosmic reionization of the IGM has been
extensively studied in the 
last years both from a theoretical and observational point of view
\citep[see][for a recent review]{Loeb2013}. While many steps towards our understanding
of reionization process have been done, still many fundamental details are missed: In
which way does it proceed? How gradual and how prolonged was the process? Was
radiation from early stars sufficient to sustain this phase
transition? Do PopIII stars or quasars have a major role in driving the
process? Is some other more exotic process at work? etc.

Historically, the exploration of the distant Universe has been
carried out following two main pathways: the observations of bright
quasars detected in wide shallow surveys \citep{Fan2012}, and of distant galaxies identified through the drop-out
technique in small fields
\citep{Bouwens2014}.  Thanks to their extreme brightness Gamma-Ray Bursts (GRBs)
represent an alternative way to access those early epochs.
As demonstrated by GRB~090423 at $z=8.2$
\citep{Salvaterra2009,Tanvir2009}, they can be detected even at
distances much larger than any other cosmic object. In principle, their afterglow emission can be
observed up to $z\sim 20$ \citep{Ciardi2000,Guo2004} providing useful
information about the ionization and metal enrichment history of the
early Universe. Here, I will review the present observational status of
this research field and discuss the possible role of GRBs in
the exploration of the Universe during and before the reionization epoch.

This paper is organized as follows. In Section~2, a brief summary of the
observations of the GRBs at $z>6$ and of their host
galaxies is given. Section~3 presents the expected rate of high-$z$ GRBs
both from PopII and PopIII stars and the expected properties
of the relative host galaxies. In Section~4, the use of GRBs as a
probe of the early
Universe is reviewed. Finally in Section~5 I present some ideas for the future of
this research field. The conclusions are drawn in Section~6.

\section{Observations}

\begin{table*}
\begin{center}
\begin{tabular}{lcccccccccc}
\hline
\hline
GRB & $z$ & $E_{\rm p}$ & $E_{\rm iso}$ & $\log(N_{\rm HI})$ & $\log(N_{\rm H,X})$ & $Z$ &
$A_V$ & $M_{\rm UV,host}$ & SFR$_{\rm host}$ \\
 & & [erg] & [erg] & [cm$^{-2}$] & [$10^{21}$ cm$^{-2}$] &
 [$\Zsun$] &  & [AB] & [$\Msun {\rm yr}^{-1}$] \\
\hline
050904 & 6.3 & 3178 & $1.24\times 10^{54}$ & 21.6
& $63^{+34}_{-29}$& $-1.6\pm 0.3$  &  $0.15\pm0.07$& $>-19.95$ &
$<4.1$ \\
080913 & 6.7 & 1008 & $7\times 10^{52}$ & 19.84 & $95^{+89}_{-77}$ & -
& $0.12\pm 0.03$ & $>-19.00$ & $<1.3$\\
090423 & 8.2 & 746 & $1.88\times 10^{53}$ &  - &
$102^{+49}_{-54}$& -  & $<0.1$ & $>-16.95$ & $<0.38$ \\
130606A & 5.9 & 2028  & $2.7\times 10^{53}$& 19.93 & $<30$ &$-1.35\pm
0.15$ & $<0.05$ & - & -\\
140515A & 6.3 & 376 & $5.1\times 10^{52}$& 18.62
&$<226$  &$<-0.8$ & $0.11\pm 0.02$ & - & -\\
\hline
090429B & 9.4 & 437 & $4.31\times 10^{52}$& - &$140\pm10$ &- & $0.10\pm 0.02$ &$>-19.65$ & $<2.4$\\
120521C & 6.0 & - & $1.9\times 10^{53}$ & - & $<60$ & - & $<0.05$ & - & -\\
120923A & 8.5 & 376 & $5.1\times 10^{52}$& - &$<720$ &- & - & - & - \\
\hline
\hline
\end{tabular}
\end{center}
\caption{List of the GRBs at $z>6$ detected by {\it Swift} and of
  their observed properties: peak energy ($E_{\rm p}$), isotropic equivalent
  energy in the 1-10000 keV range ($E_{\rm iso}$), hydrogen column
  density in the host ($N_{\rm HI}$), X-ray equivalent hydrogen column
  density ($N_{\rm H,X}$), host metallicity ($Z$) and dust extinction ($A_V$). The last two columns report
  the limits on the host galaxy luminosity ($M_{\rm
    UV,host}$) and SFR. See text for references. }
\end{table*}

\subsection{The Swift high-$z$ GRB sample}

In ten years of operations {\it Swift} has detected a handful of bursts
with spectroscopically confirmed redshift larger than 6. In
addition, other three GRBs have well constrained photometric redshift above
this limit. The
observed high-$z$ sample represents $\sim 1$\%  of all {\it 
  Swift} bursts, $\sim 2.5$\% of those with known $z$. The main properties of the $z>6$ GRB sample
are given in Table~1.

\begin{itemize}
\item{{\bf GRB~050904} at z=6.3 \citet{Kawai2006}}

This burst was firstly imagined by the 25-cm telescope TAROT \citep{Klotz2005}. Its high-$z$ nature was recognized by multi-wavelength
photometric data \citep{Haislip2006,Tagliaferri2005} and firmly confirmed
spectrocopically three days after the {\it Swift} trigger by the
Subaru telescope \citep{Kawai2006}. The afterglow spectrum provided
a upper limit on the neutral hydrogen fraction at the GRB
redshift of $x_{\rm HI}<0.17$ at 1$\sigma$ confidence \citep{Totani2006} and a measure of the metallicity at the
level of $\sim 0.1\;\Zsun$ \citep{Kawai2006}. Recently, \citet{Thoene2013}
revised this value, inferring a slightly lower metallicity from the S
II $\lambda1243$ equivalent width, resulting in $\log(Z/\Zsun)=-1.6\pm 0.3$. The afterglow spectral energy
distribution (SED) requires the presence of SMC or supernova (SN) type
dust at a level of $A_V=0.15\pm 0.07$ (\citealt{Stratta2011} but see
\citealt{Zafar2011a}). The modeling of afterglow data from X-ray to
radio suggests GRB~050904 to be an energetic burst blowing up in a dense
medium  with $n\simeq 680$ cm$^{-3}$ \citep{Frail2006,Laskar2014}.

\item{{\bf GRB~080913} at $z=6.7$ \citep{Greiner2009}}

 The high-$z$ nature of this burst was recognized via the detection of
 a spectral break between the i$^\prime$ and z$^\prime$ bands of the
 GROND instrument and then confirmed spectroscopically by VLT observations. The analysis of the red damping wing constrained $x_{\rm HI}<0.73$ at
 90\% confidence level \citep{Patel2010}.  In spite of the rapid
 follow-up campaign, the faintness
 of the afterglow prevented the detection of any metal absorption line, but
 the S$_{\rm II}$+Si$_{\rm II}$ at 2.9$\sigma$ level \citep{Patel2010}.
 Dust absorption of  $A_V=0.12\pm 0.03$ is found from SED fitting \citep{Zafar2011b}.

\item{{\bf GRB~090423} at $z=8.2$ \citep{Salvaterra2009,Tanvir2009}}

The spectroscopic redshift of this burst was secured by TNG \citep{Salvaterra2009} and by VLT
\citep{Tanvir2009}, and still represents the distance record for a
cosmic object. Radio observations by
VLA were reported by \citet{Chandra2010}. The analysis of the
multi-wavelength dataset show that the afterglow is reminiscent of many
other lower redshift bursts, suggesting that in spite of its extreme
redshift, its progenitor and the medium in which it blowed up were not
peculiar. Indeed, its detection is consistent with the
high-$z$ tail of PopII/I GRB redshift distribution
\citep{Salvaterra2009}. From the X-ray to optical SED no absorption by dust is
evident with $A_V<0.1$  (\citealt{Salvaterra2009,Tanvir2009}, but see \citealt{Laskar2014}). 

\item{{\bf GRB 130606A} at $z=5.9$
    \citep{Chornock2013,CastroTirado2013,Totani2014,Hartoog2014}}

The redshift of this burst was obtained by Gemini-North
\citep{Chornock2013}, GTC \citep{CastroTirado2013}, Subaru \citep{Totani2014} and VLT \citep{Hartoog2014}. In particular, the
superior resolution and wavelength coverage of the VLT/X-shooter
instrument showed the potentiality of GRBs as tool to study in great
details the metal enrichment of star forming region inside high$-z$ galaxies
\citep{Hartoog2014}. Precise column densities of H, Al, Si and Fe are
reported together with limit on C, O, S and Ni. The host metallicity
is constrained to be in  the range of 0.03-0.06 solar and the high [Si/Fe] in
the host suggests the presence of dust depletion (though $A_V<0.05$
from SED fitting). The best fit of
the Ly$\alpha$ absorption line is obtained for $\log(N_{\rm
  HI})=19.94\pm 0.01$ and negligible neutral hydrogen in the external
medium, with $x_{\rm HI}<0.03$ at 3$\sigma$ significance.

\item{{\bf GRB 140515A} \citep{Chornock2014,Melandri2015}} 

The redshift of this burst has been secured by Gemini-North
\citep{Chornock2014}, GTC and VLT
\citep{Melandri2015}. \citet{Chornock2014} analysed the Gemini-North
spectra finding no evidence
of narrow absorption lines, indicating a host metallicity
$Z<0.15\;\Zsun$. However, \citet{Melandri2015} by modeling the X-ray to optical SED found evidence for
dust absorption to the level of $A_V\sim 0.1$ indicating some metal enrichment. 
The red damping wing of Lyman-$\alpha$ can be fitted
equally well by a single host galaxy absorber with $\log(N_{\rm
  HI})=18.62\pm 0.08$ or a pure IGM absorption with neutral hydrogen
fraction $x_{\rm HI}\sim 0.06$ \citep{Chornock2014}. 
\end{itemize}

Other three GRBs have accurate photometric redshift measurement that place
in the $z>6$ sample:

\begin{itemize}
\item{{\bf GRB 090429B} at $z\sim 9.4$ \citep{Cucchiara2011}}

\citet{Cucchiara2011} collected the afterglow data obtained with
Gemini-North, VLT and GROND. In the best
  fit model these data are all consistent with a photometric redshift of $z=9.4$ and low
  extinction $A_V=0.10\pm 0.02$. A secondary solution at very low
  redshift is still allowed by the SED fitting, but it seems
  unlikely due to the lack of any
  detection of the GRB host galaxy (see next Section).

\item{{\bf GRB 120521C} at $z\sim 6$ \citep{Laskar2014}} 

The large multi-wavelength dataset of the afterglows lead to
$z\simeq 6.0$ also supported by the analysis of the very low
  signal-to-noise Gemini-North spectrum. All afterglow
  data after $\sim 0.25$ days can be well fitted assuming
  constant-density medium with  $n_0\sim 0.05$ cm$^{-3}$ and no dust
  extinction, $A_V<0.05$. The radio observations revealed the existence
  of a jet break corresponding to an jet angle $\theta_{\rm jet}\simeq 3$
  degrees. An  isotropic energy $E_{\rm iso}=1.9\pm0.8 \times 10^{53}$
  erg is obtained by extrapolating  the BAT fluence.

\item{{\bf GRB 120923A} at $z\sim 8.5$ \citep{Tanvir2013}}

The optical-NIR afterglow of this burst was imagined by Gemini-North
telescope $\sim 1.4$ hours after the {\it Swift} trigger. The blue IR color, H-K$\sim 0.1$ mag together
with the break between the Y- and J-band is suggestive of a strong
high-$z$ candidate \citep{Levan2012}. The fit to all photometric data gives $z\sim
8.5$ \citep{Tanvir2013}.  No radio or mm 
detection is reported. 
\end{itemize}

\subsection{General properties of high-$z$ GRBs}\label{sec:obsprop}

In spite of the burst to burst differences and of the small size of
the sample, some interesting clues about the nature
of high-$z$ bursts can be gathered by comparing them with  
bursts at lower redshifts. Thus, in the following, I will compare the
properties of $z>6$ bursts with those of a well selected,
complete sample of bright {\it Swift} bursts \citep[BAT6,][]{Salvaterra2012}. The BAT6 sample has been selected only on the basis
of the $\gamma$-ray peak flux as seen by {\it Swift}/BAT and is
characterized by a very high completeness in redshift \citep[95\%,][]{Salvaterra2012,Covino2013}. Thus, it represents a perfect sub-sample to
compare the results of the high-$z$ burst population with.

The top-left panel of Fig.~1 shows the position of $z>6$ bursts in the
$E_{\rm p}-E_{\rm iso}$ plane. The black points represents the BAT6 sample
\citep{Nava2012} and the shaded area shows the 3$\sigma$ scatter around the
$E_{\rm p}-E_{\rm iso}$ correlation. 
High-$z$ bursts nicely follow the Amati correlation suggesting that their
prompt emission properties do not differ significantly from bright
lower-$z$ bursts. This consistency calls for a
common central engine (and progenitor star).

\begin{figure*}\label{fig1}
\centering
 {\includegraphics[scale=0.4]{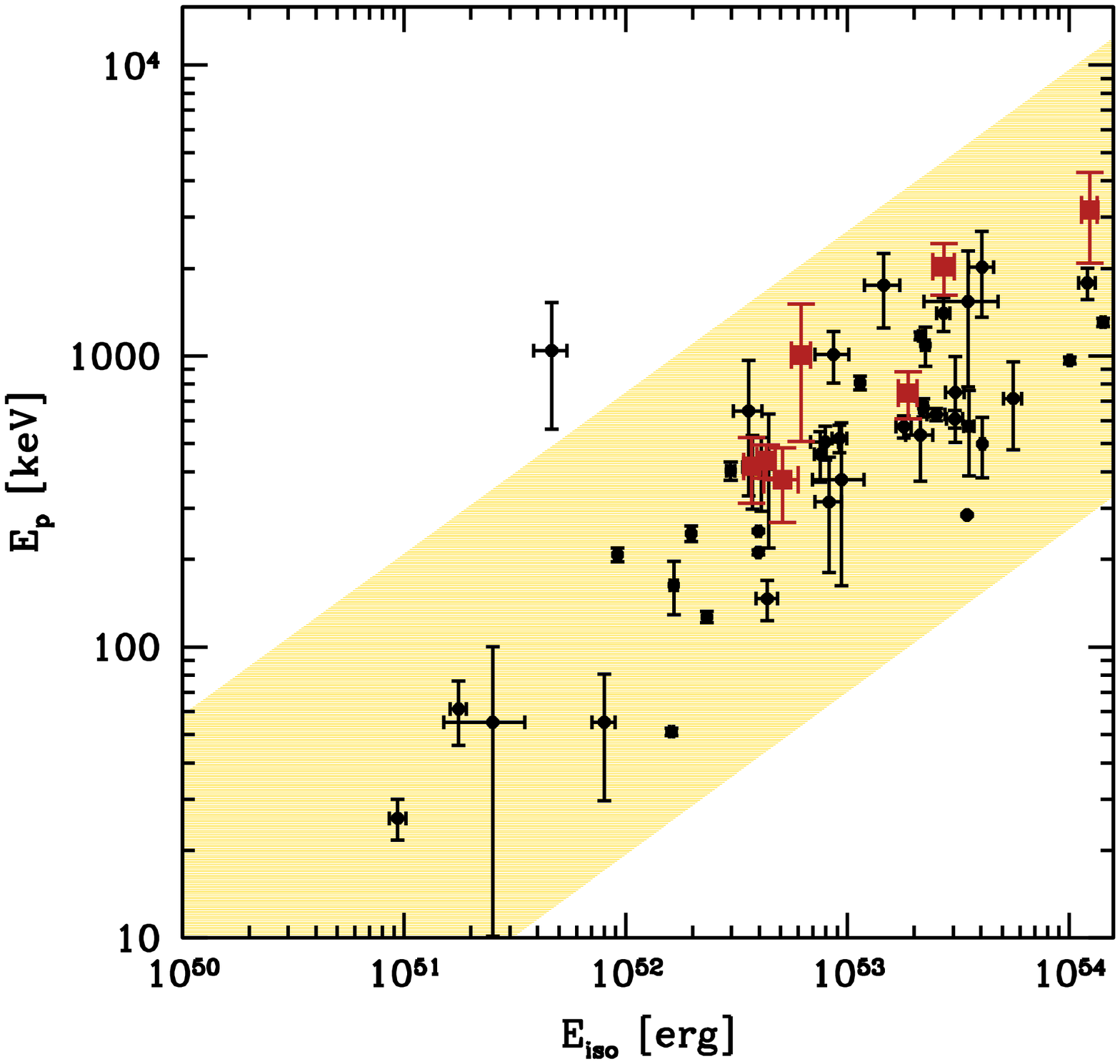}}
 {\includegraphics[scale=0.4]{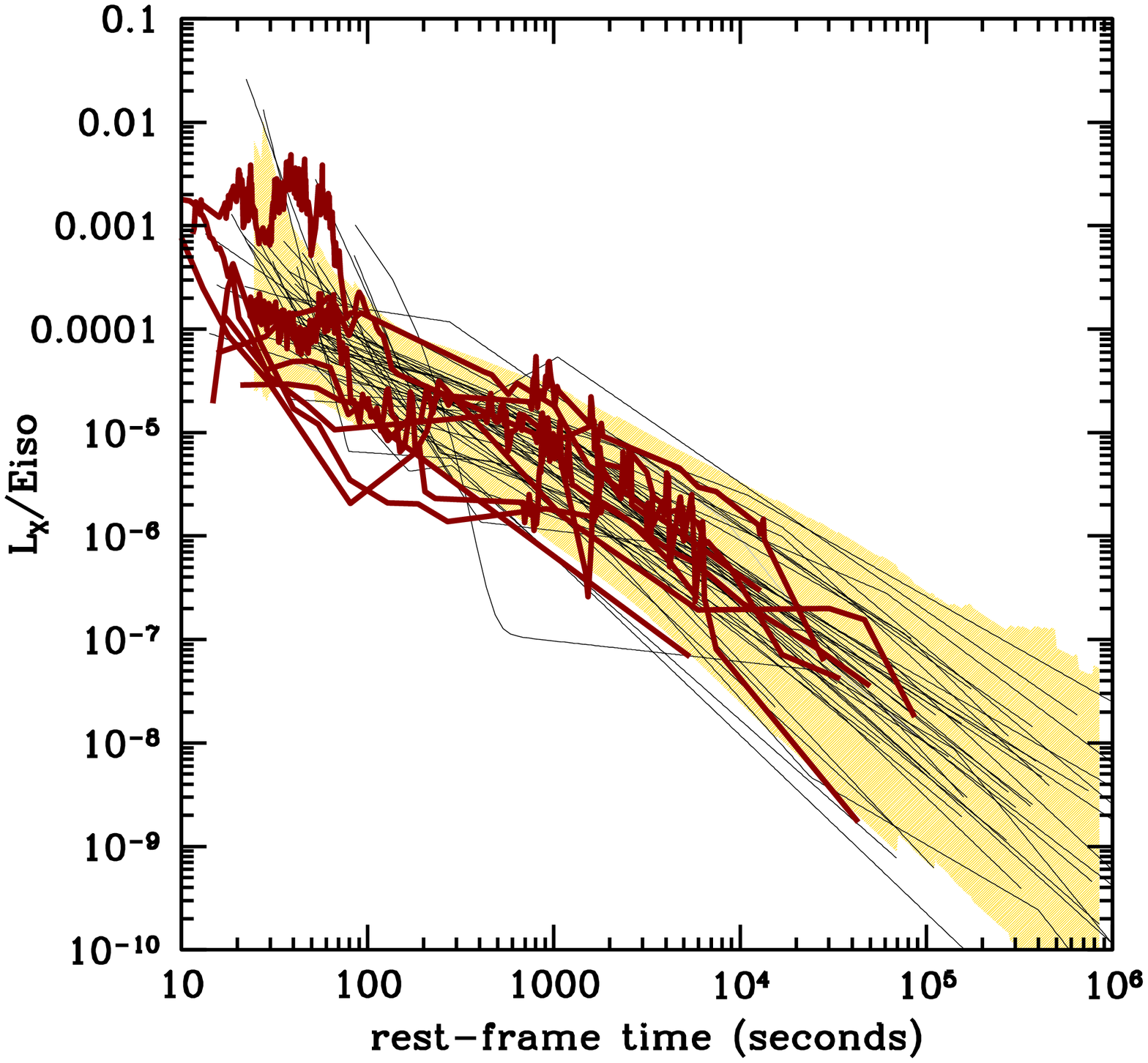}}
{\includegraphics[scale=0.4]{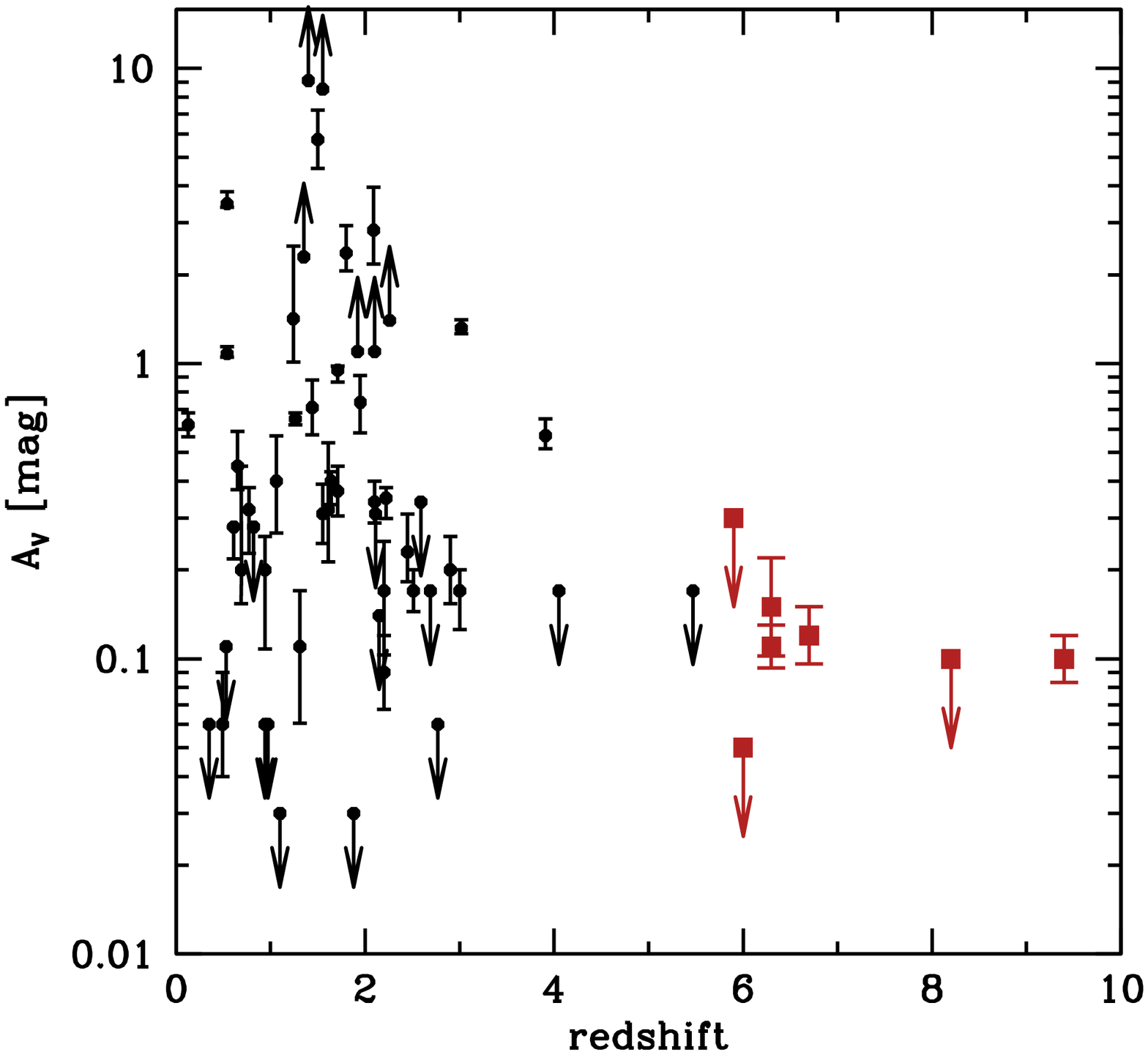}}
 {\includegraphics[scale=0.4]{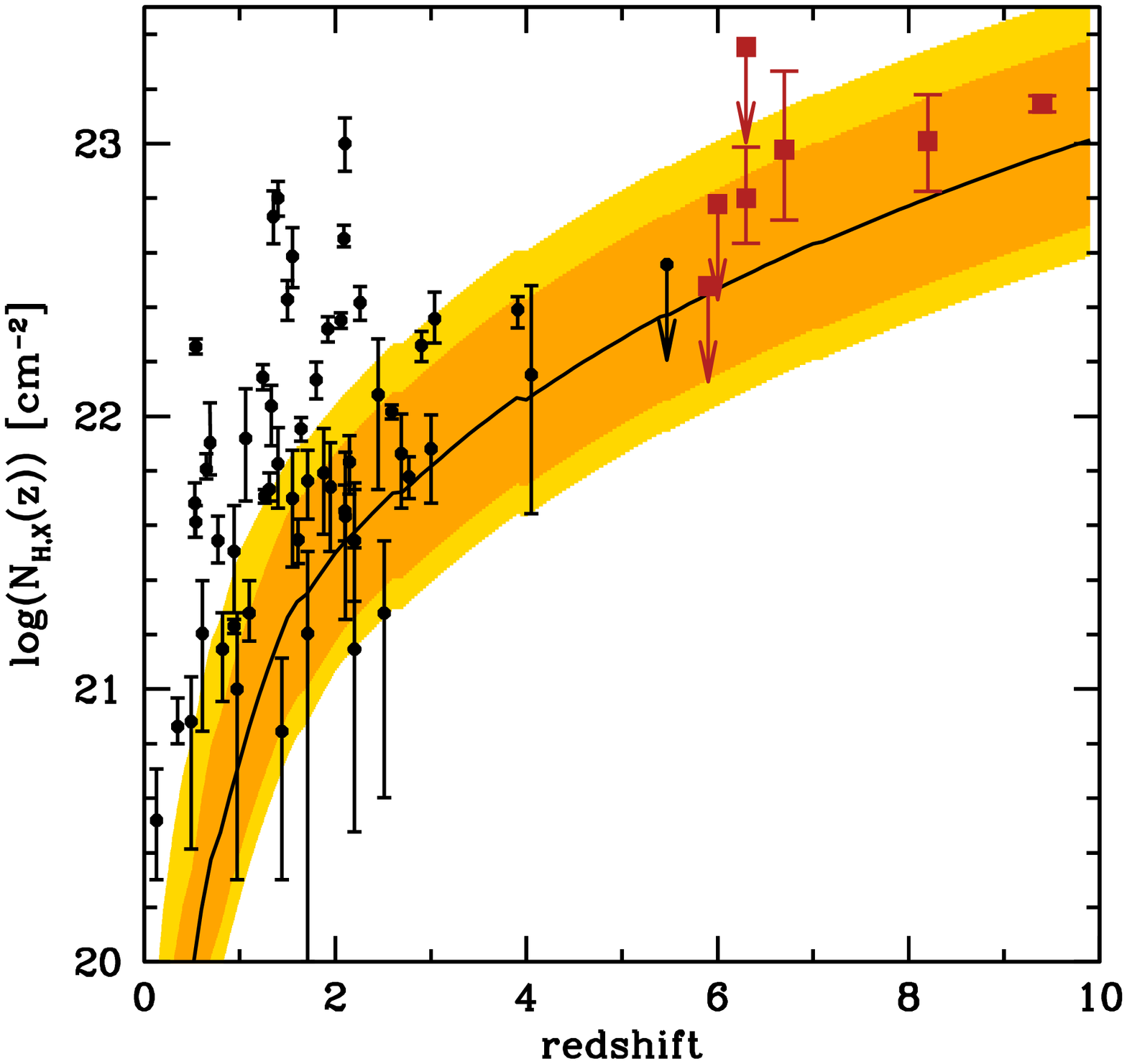}}
\caption{Comparison of the properties of $z>6$ bursts (red squares and
  lines) with those of a
  well selected complete sub-sample of bright {\it Swift} GRB, the BAT6
  sample (black points and lines; \citealt{Salvaterra2012}). {\it
    Top-left panel}: $E_{\rm p}-E_{\rm iso}$ correlation where BAT6 bursts are
  from \citet{Nava2012}. The shaded regions represents the $3\sigma$ scatter around the
  best-fitting relation. {\it Top-right panel}:
rest-frame 2-10 keV light curves normalized to their isotropic energy where BAT6
  data are from \citet{DAvanzo2012}. The shaded region 
  represents the 2$\sigma$ scatter around the mean value of
  the $L_{\rm X}/E_{\rm iso}$ distributions at the given rest-frame time. {\it Bottom-left panel}: dust extinction as a
  function of redshift, where BAT6 data are from \citet{Covino2013}. {\it Bottom-right panel}: 
  intrinsic X-ray equivalent hydrogen column densities $N_{\rm H,X}$ as
  a function of  redshift, where BAT6 data are from \citet{Campana2012}. The shaded regions represent the effect of intervening
  material along the line-of-sight (see \citealt{Campana2015} for the
  details).}
\end{figure*}

In the top-right panel of Fig.~1 I have then compared the rest-frame
2-10 keV light
curves normalized for the corresponding $E_{\rm iso}$ of $z>6$
bursts with those obtained from the BAT6 sample \citep{DAvanzo2012}.
Indeed, the X-ray light curves of low-$z$ GRBs tend to cluster when such a
normalization is performed and strong correlation between $L_{X}$ and
$E_{\rm iso}$ is found at early times. This effect is seen also in high-$z$
GRBs and their X-ray light curves can not be distinguish from 
those of low- and intermediate-$z$ bursts. This conclusion is also supported by more extensive,
although model dependent, analysis of afterglow emission. It has been shown
\citep{Chandra2010,Laskar2014} that the total energy,
microphysical parameters and medium properties of the high-$z$
GRBs are in line with those derived for lower redshift bursts.

In the bottom-left panel of Fig.~1, the available host dust extinction
estimates for $z>6$ bursts are plotted against those obtained for the
BAT6 sample \citep{Covino2013}. Being selected only on the
basis of the GRB $\gamma-$ray peak flux, the BAT6 sample is representative of the
extinction distribution in GRB selected galaxies: 
50\% of bursts suffer less than 0.3-0.4 mag extinction and
only 13\% of GRBs have $A_V>2$
mag. The high-$z$ sample is characterized by little or no
absorption\footnote{However, the lack of highly
extincted bursts may reflect a bias in the sample
as even a small amount of absorption at $z>6$ will prevent the
detection of the optical afterglow and thus of the redshift measure.}, 
suggesting a decrease in dust content in
star-forming environments at high redshifts \citep{Zafar2011b}. Similarly, the metallicities inferred for $z\sim 6$ bursts are in line,
though at the lower end, with the distribution of $Z$ measured in
lower redshift GRB afterglows \citep{Savaglio2009,Sparre2014}. The observed $Z$ and $A_V$ values (or limits) are indeed in agreement
with those expected for high-$z$ galaxies populating the
faint end of the luminosity function \citep[see e.g.][]{Salvaterra2011}.

At variance with lower redshift bursts, all high-$z$ GRBs have measured (or limit consistent with) very high
X-ray equivalent hydrogen column densities, $N_{\rm H,X}$ (bottom-right
panel of Fig.~1). At lower
redshift, a high $N_{\rm H,X}$ is usually connected to highly extincted afterglows \citep{Campana2012,Covino2013}.
However, at higher redshifts this interpretation seems to be unrealistic
given our knowledge of galaxy evolution at early times. Moreover, little or no dust absorption is 
found in high-$z$ bursts, calling for unusual metal-to-dust ratios.
In recent years, a well defined trend of increasing $N_{\rm
  H,X}$ with redshift has been found using both GRBs and AGNs
\citep{Campana2010,Behar2011,Starling2013,Campana2015}.  High-$z$ GRBs extend this
relation to higher redshifts quite nicely with no $N_{\rm H,X}$ value
or limit below the observed lower envelope.
\citet{Campana2015} have shown, by means of dedicated numerical simulations, that this effect can be naturally
explained by the 
absorption of intervening metals along the line-of-sight (LOS), being
the observed lower
envelope due to metals present in the IGM. An additional
contribution in most of the LOSs comes from metals in the diffuse
gas in galaxy groups at $z\sim 0-2$.

In conclusion, high-$z$ GRBs do not show any peculiar feature with
respect to low- and intermediate-$z$ events. Both the energetics and
the afterglow properties are found to be very similar. The medium
in which they blow up does not differ too much in terms of density, metallicity and
dust content. Finally, also the duration of their prompt emission is
consistent with those of the low-$z$ sample taking into account
cosmological and instrumental effects \citep{Littlejohns2013}. All these findings support the idea that the GRBs
detected so far at $z>6$ represent the high redshift tail of the PopII/I GRB redshift distribution.

\subsection{High-$z$ GRB host galaxies}\label{sec:obshost}

In the last few years, deep searches with both ground- and space-based
facilities \citep{Tanvir2012,Basa2012,Walter2012,Berger2014} have been
carried out searching for the host galaxy of 
high-$z$ GRBs. Indeed, the a-priori, precise knowledge of the position and
distance of the galaxy provided by the GRB detection allows to 
optimize the telescope setup, pushing the instruments to their
limit.

\citet{Tanvir2012} reported the observation of the field of four
GRBs at $z>6$ with HST with filters choosen to cover the rest-frame wavelength range
1200-1500\AA, i.e. above the Lyman-$\alpha$. In spite of the very deep
limits reached (see Table~1), none of the targets
have been identified. Similar results were obtained by \citet{Basa2012} using slightly shallower VLT observations.
More recently, \citet{Berger2014} imagined the field
of GRB~090423 with {\it Spitzer} at 3.6 $\mu$m and ALMA at 222 GHz,
but, also in this case, no detection is reported. These results imply that all the
hosts lie below the characteristic luminosity value at their respective redshifts, with star
formation rates SFR$<4\;\Msun$ yr$^{-1}$ in all cases. In particular, GRB~090423
was possible to derive a strong limit on the unobscured
SFR$<0.38\;\Msun$ yr$^{-1}$ \citep{Tanvir2012}, while the ALMA non
detection required the obscured SFR to be $<5\;\Msun$ yr$^{-1}$ \citep{Berger2014}. 

\citet{Tanvir2012} also stacked all images deriving a limit on the mean
SFR per galaxy of $<0.17\;\Msun$ yr$^{-1}$, consistent with the idea
that the bulk of star formation activity is missed in current deep HST
surveys \citep{Salvaterra2011}. Moreover, these findings offer independent
evidence that the galaxy
luminosity function is evolving rapidly with redshift with a
steeper faint-end slope or with a decreasing characteristic
luminosity at high redshift \citep{Tanvir2012}.

\section{Theory}

\subsection{The high-$z$ GRB population}

As described in previous Section, {\it Swift} has detected almost one
GRB at $z>6$ per year. Considering the 1.4 sr field of view of the BAT
instrument, this corresponds to a rate\footnote{While this represents a lower limit, given the fact that
some high-$z$ burst can be among GRBs without redshift measurement, it
should be noted that there is a possible positive bias at work here. Indeed,
most of the observational programs at large ground- and
space-facilities have as their primary goal the detection of high-$z$
events.} of $\sim 0.6$ bursts yr$^{-1}$
sr$^{-1}$. It is worth to ask whether such a rate is expected on the
basis of our knowledge of the intrinsic GRB redshift distribution and luminosity
function \citep{Bromm2002,Daigne2006,SalvaterraChincarini2007,Salvaterra2009b,Butler2010,Wanderman2010,Robertson2012,Salvaterra2012,Ghirlanda2015}.

\begin{figure}\label{fig:ngt6}
\centering
 {\includegraphics[scale=0.38]{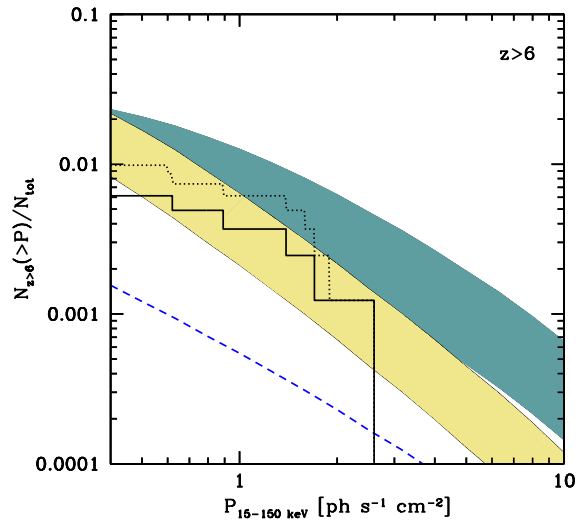}}
\caption{Peak flux distribution of bursts at $z>6$ as measured in the
  15-150 keV {\it Swift}/BAT band. The solid (dotted) histogram shows the distribution of
  bursts with spectroscopic (spectroscopic and photometric) redshift. The
  shaded regions report model predictions \citep{Salvaterra2012}, taking into account the uncertainties in the
  determination of the GRB luminosity function and evolution. Dark (light) color refers to the density
  (luminosity) evolution models. The dashed line is obtained assuming
  no evolution and clearly
  underestimates the number of high-$z$ detections at all peak fluxes.}
\end{figure}

Figure~2 shows the distribution of peak
fluxes of bursts at $z>6$ as measured in the 15-150 keV {\it
  Swift}/BAT band. The histogram is compared with the model results under
different assumptions on the evolution of the GRB luminosity function
\citep{Salvaterra2012}. The models have
been calibrated by jointly fitting the differential peak
flux distribution of BATSE long GRBs and the redshift distribution of
the BAT6 sample. The models reproduce the rate and peak flux
distribution of $z>6$ burst remarkably well\footnote{Since the BAT6 sample is limited to
$z<5.4$ there is no assumption in the model about the peak flux distribution of
$z>6$ GRBs}
with a small excess in the predicted population at the faintest
fluxes. However, since, in general, to faint
$\gamma$-ray fluxes correspond faint optical afterglows
\citep[e.g.][]{Melandri2014}, some high-$z$ faint bursts can be missed
being the detection of its afterglow difficult even with 8-m class telescopes. 

By extrapolating model results to
fainter fluxes \citep{Salvaterra2008,Ghirlanda2015}, it can be shown that bursts $z>6$ represent $\sim 10$\% of the whole population,
suggesting that GRBs are quite efficient in selecting high-$z$ objects. The detection of
these events is one of the main goals of any future GRB
mission.

\subsection{GRBs from PopIII stars}\label{sec:popIII}

Different authors \citep{Meszaros2010,Komissarov2010,Suwa2011,Toma2011,Nagakura2012,Nakauchi2012,Piro2014} have proposed
that the conditions for jet breakout could be met also during the
final phases of the collapse of a PopIII massive star.  Although under
different frameworks, all models consistently predict PopIII GRBs to be
very energetic events, with total energies
exceeding by orders of magnitude those expected in PopII events. In particular
\citet{Toma2011} suggested that PopIII bursts could
have an equivalent isotropic energy $\sim 10^{56-57}$ erg making their
detection possible even at the highest redshifts. A much longer
prompt emission, with typical duration of $10^4$ s, is also foreseen
in most of the models.
 However, all these
characteristics are shared  observationally with the population of ultra-long GRBs
recently detected at much lower redshift \citep{Levan2014} and likely
associated with PopII
blue supergiant stars \citep{Nakauchi2013}. Thus, they do not represent a unique feature to
firmly indentify a PopIII event. Strong evidence 
could be provided by the absence of metal absorption lines in the
afterglow spectrum down to a level of the critical
metallicity. However, the measure of such low metallicity (or a limit
consistent with it) requires extremely high signal-to-noise spectra
that can be even beyond the capabilities of
30-m class telescopes. In the absence of any other observational tool
to identify PopIII events, radio observations can provide an important
clue. Indeed extremely powerful radio afterglows
\citep{Ciardi2000,Toma2011,Ghirlanda2013b}  are expected given the peculiar energetics of PopIII
events. \citet{Ghirlanda2013b} showed that PopIII radio
afterglows should reside in a very distinct position in observed  peak
time-peak flux plane, reaching much larger peak fluxes at later times
than any PopII GRBs.
\smallskip

We have seen in Section~\ref{sec:obsprop} that it is unlikely that the observed $z>6$
bursts are associated with PopIII progenitors. This piece of
information has been used to put a limit on the expected rate of
PopIII events \citep{Bromm2006,Campisi2011,Toma2011,deSouza2011,MaioBarkov2014,Mesler2014}. By means of dedicated numerical simulations including the
effect of the chemical feedback, \citet{Campisi2011} computed the
expected redshift distribution of PopIII GRBs. The PopIII GRB rate is
found to 
increase from $z=20$ to $z=8-10$ and then it decreases rapidly becoming
negligible at $z\sim 
5$. Under the assumption that all PopIII GRBs can be detected by {\it
    Swift}\footnote{This may be not the case since the {\it Swift} trigger algorithm \citep{Lien2014} is not very sensitive to extremely long GRBs.}, the maximum allowed
rate is $\sim 0.03$ Gpc$^{-3}$ yr$^{-1}$. Assuming a typical jet angle
of $\sim 7$ degree and a typical PopIII mass of $\sim 100\;\Msun$,
this corresponds to $<1$ PopIII GRB every 500 PopIII stars, comparable
with the PopII/SN type Ib/c ratio \citep{Ghirlanda2013a}. 

\medskip

In conclusion, even if GRBs can really result from the 
collapse of a massive PopIII star, they will be extremely
rare with $<0.05$ event yr$^{-1}$ sr$^{-1}$ at $z>6$. PopIII GRBs
become the dominant observed population only at $z>10-15$
\citep{Campisi2011}. On the other hand, even a single detection of a
GRB powered by PopIII stars, i.e. a extremely high-$z$,
very energetic, long event with a bright radio afterglow peaking a
late times, would represent a breakthrough in the study of the first
generation of stars. 

\begin{figure}
\centering
 {\includegraphics[scale=0.35]{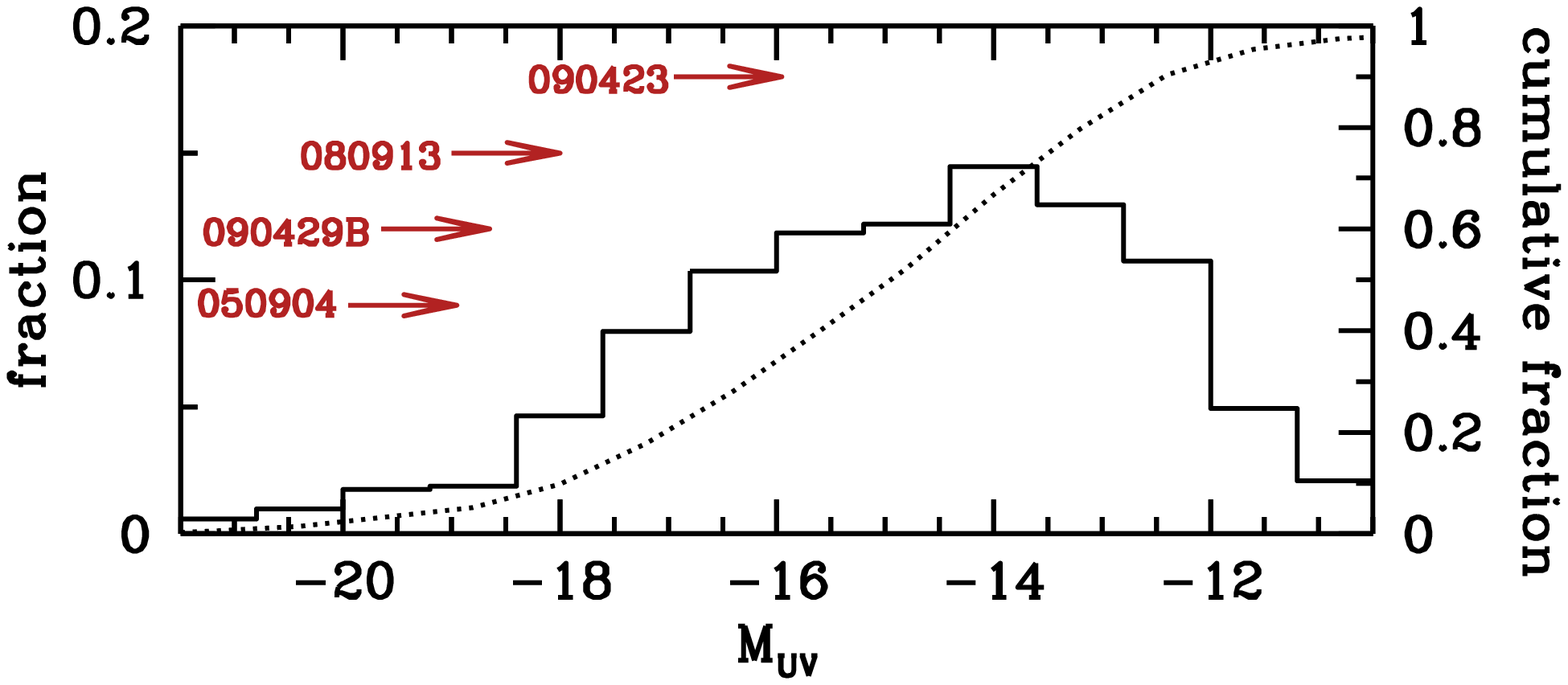}}
{\includegraphics[scale=0.35]{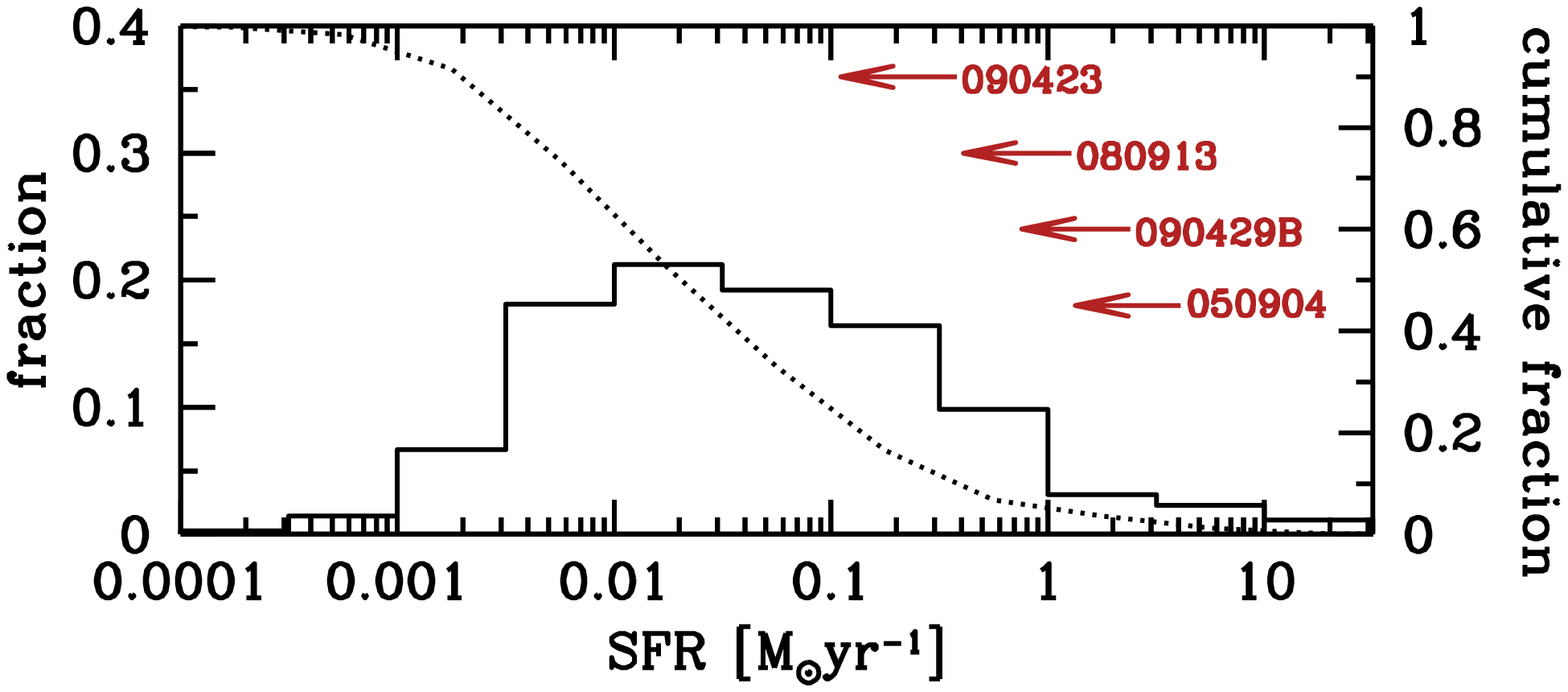}}
  {\includegraphics[scale=0.35]{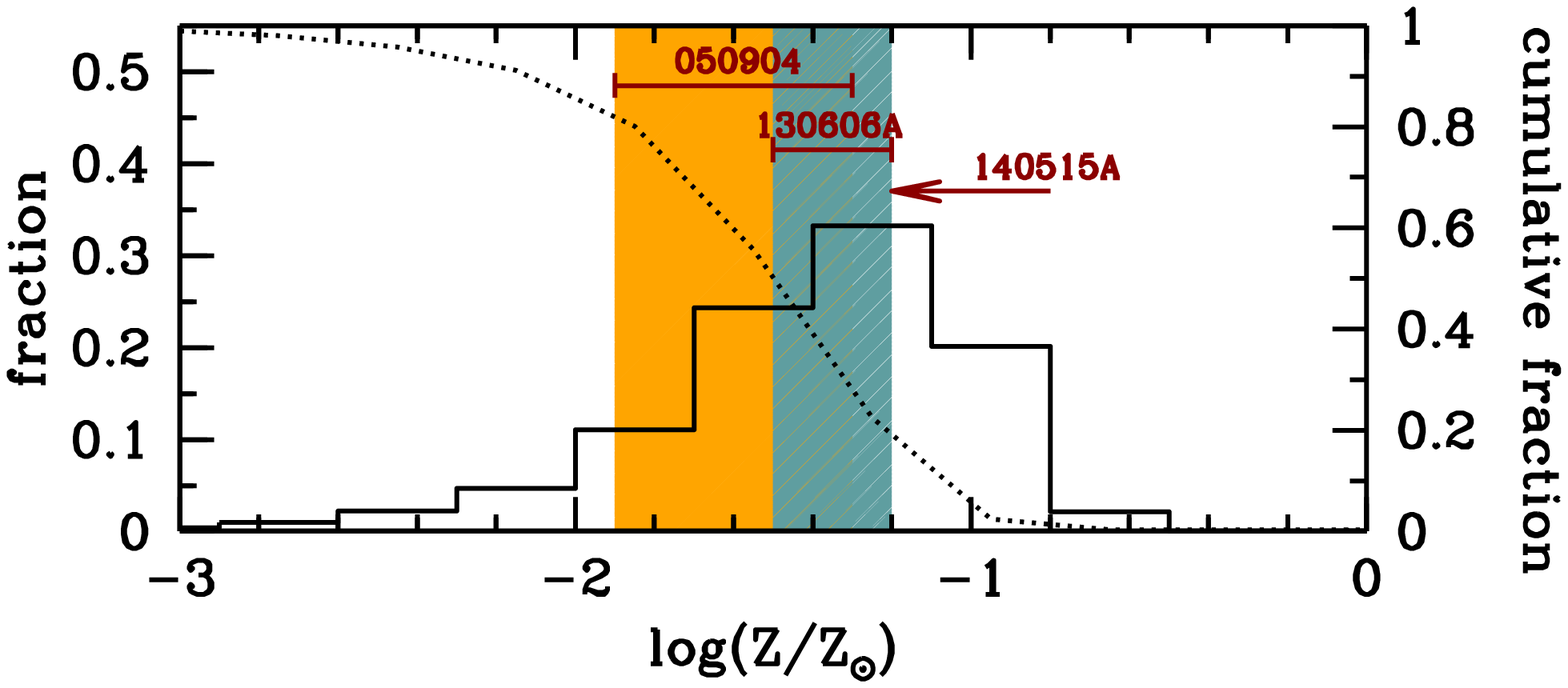}}
\caption{
Properties of simulated GRB hosts in the redshift range
$z=6-10$. Panels from top to bottom show the distribution of UV
absolute magnitude, of the SFR (in
$\rm \Msun~yr^{-1}$), and of gas phase metallicities (in solar units). The dotted line shows the normalized cumulative
distributions (see right y-axis). The arrows in the first two panels
report the limits on $M_{\rm UV}$ and SFR of GRB hosts \citep{Tanvir2012}.
The shaded areas in the bottom panel correspond to the metallicity measured
in the GRB~050904 \citep{Thoene2013} and GRB~130606A
\citep{Hartoog2014} afterglow spectrum. The
arrow shows the limit inferred  for GRB~140515A \citep{Chornock2014}.
}
\end{figure}

\subsection{High-$z$ GRB host galaxies}

In absence
of any detection (see Section~\ref{sec:obshost}), the
relation of GRB selected galaxies with the typical high-$z$ galaxy
population has been studied by means of dedicated
semi-analytical models \citep{Trenti2012} or numerical
simulations \citep{Salvaterra2013,Elliott2015}.

\citet{Salvaterra2013} derived the expected properties
of the GRB hosts at $z=6-10$ by means of cosmological numerical simulations including all
relevant feedback effects at these redshifts. GRBs are found to
explode in bursty galaxies with typical star formation rates
SFR$\simeq 0.03-0.3$ M$_\odot$ yr$^{-1}$, stellar masses
$M_\star\simeq 10^6-10^8$ M$_\odot$, specific star formation rates
sSFR$\simeq 3-10$ Gyr$^{-1}$.  The distribution of their UV
luminosities (see top panel of Fig.~3) places them in the faint end of the galaxy
luminosity function, below the capabilities of current
instruments. This is consistent with the lack of detection and suggests that deep observations with the
 James Webb Space Telescope (JWST) are required to pinpoint them. Finally, the bulk of the GRBs
are found to explode in galaxies already enriched with metal to
$Z\simeq 0.03-0.1$ Z$_\odot$ (bottom panel of Fig.~3), consistently with the
metallicity inferred in $z\sim 6$ GRB afterglow spectra
(\citealt{Salvaterra2013}, but see \citealt{Cen2014}).

The same simulations allow to predict the properties of host galaxies
of PopIII GRB events \citep{Campisi2011}. These bursts are
found to reside typically in objects at the lower end of the stellar
mass with $M_\star< 10^7\;\Msun$ as they have the
higher probability of having still a pristine composition. While a
metallicity below $Z_{\rm crit}$ is obviously required in the star
forming cloud from which the PopIII GRB arises, some of the PopIII
GRB hosts are found in galaxies with $Z\sim 10^{-3}\;\Zsun$. This may
reflect the fact that some PopIII event can be associated with 
pockets of metal-free gas in the outskirt of a enriched galaxy
\citep{Tornatore2007}.

\section{High-$z$ GRBs as a tool}

\subsection{Independent measure of the SFR}

It is well established that long GRBs are associated to the death of
massive stars \citep[see][and references
therein]{Hjorth2012}. Therefore, they should trace to some extent the star
formation activity through cosmic times and they can be use to measure
in an independent way the global SFR
\citep{Kistler2009,Ishida2011,Robertson2012}. In this respect, GRBs
have many advantages compared with usual probes: (i) they are detected at higher
redshifts; (ii) they are independent on the galaxy brightness; (iii)
they do not suffer of usual biases affecting optical/NIR surveys. 
Indeed, although hampered by small statistics, the
detection of the few high-$z$ GRBs already suggests that the global
SFR at $z\simeq 8-9$ is 3-5
times higher than deduced from high-$z$ galaxy searches through
the drop-out technique \citep{Ishida2011}.
An important caveat to be mention here is that GRBs may  be not perfect
tracers of the SFR. Indeed, in order to reproduce the redshift
distribution of bursts detected by {\it Swift}
\citep[e.g.][]{SalvaterraChincarini2007,Wanderman2010,Robertson2012,Salvaterra2012}, some kind of evolution with redshift is required. Thus, a precise
knowledge of the nature and the value of this bias 
is needed to properly use GRBs as SFR probe in the early
Universe. This can be achieved by measuring the GRB LF and
its evolution at low and intermediate redshifts
\citep{Salvaterra2012,Wanderman2010,Pescalli2015} and by studying well selected sample of low$-z$ GRB host
galaxies
\citep{Boissier2013,Perley2013,Hunt2014,Trenti2014,Vergani2014}. In
particular, \citet{Vergani2014} have shown that the stellar mass
distribution of a small, but complete, sample of GRB selected galaxies at $z<1$ is consistent
with the existence of a mild metallicity bias in the GRB hosts, with
bursts forming more efficiently in environments with
$Z<0.3-0.5\;\Zsun$ \citep{Vergani2014}. If these findings are
confirmed by larger samples, GRBs can be considered good tracers of
the SFR at least at $z>3-4$.

\subsection{The reionization history}

Similar to quasars, GRBs have been used to measure the neutral hydrogen
fraction in the IGM by fitting the red damping wing of the Ly$\alpha$
and therefore to place strong constraints on the reionization redshift
\citep{Totani2006,Chornock2013,Totani2014,Hartoog2014,Chornock2014}. 
With respect to quasars, GRBs have many advantages: (i) they are already
detected at much larger redshift than quasars; (ii) they reside in
average cosmic regions, less affected by local ionization effects or
clustering, and (iii) their spectrum follows a power-law making
continuum determination much easier. The main limitation of this
method is represented by the presence of the host galaxy absorption
\citep{McQuinn2008}, although this can be in principle separated being
less extended. \citet{Chen2007} found that $\sim 20-30$\% of GRBs have small enough hydrogen column
density to allow a direct measure of absorption from a partially
neutral IGM. Moreover, the typical $N_{\rm HI}$ value for GRB hosts is
expected to decrease with increasing redshift \citep{Nagamine2008}.
This idea is now supported by the fact that three out of the four $z>6$ GRBs for
which the intrinsic HI column density has been measured, showed
$N_{\rm HI}<10^{20}$ cm$^{-2}$ (see Table~1). In particular, a very
low $N_{\rm HI}$ is found in the case of GRB~140515A
\citep{Chornock2014} suggesting that a statistical sample of bursts
with low intrinsic
column densities can be gathered in future allowing to firmly
constrain the reionization process.

Another effective method to constrain the reionization history  is
based on the statistics of peaks/gaps in the afterglow spectrum between Ly$\alpha$
and Ly$\beta$, i.e. corresponding to transmission/absorption regions along
the LOS \citep{Paschos2005}. \citet{Gallerani2008} have shown that
the distribution of the dimension of the largest dark gap is sensitive
to the assumed reionization history 
provided that a statistical sample of $\sim 20$ LOSs is gathered.

Finally, the study of the HI forest over-imposed on the GRB radio
afterglow with SKA of a very high redshift burst could offer, in principle, a powerful tool to
study the reionization process \citep{Xu2011,Ciardi2015}. However, it requires
the detection of an extremely bright radio afterglow at the level of
$\sim 10$ mJy at $z>7-8$, that may be possible in the case of PopIII GRBs.

Beside constraining the reionization history and its variance along
different LOSs, GRBs can provide an estimate of the typical escape
fraction of UV photons from early galaxies. The measure of this
crucial quantity from the observation of the ecaping
Lyman continuum radiation is already very difficult at $z\sim 3$
\citep{Vanzella2012}, and may be
impossible for the small, faint galaxies responsible for the bulk of star
formation at $z>6$. A statistical sample of GRBs provide an
alternative way to infer it from
the distribution of $N_{\rm HI}$ over many LOSs.  Useful
constrains have so far obtained at $z=2-4$ with $f_{\rm esc}<7.5$\% \citep{Fynbo2009}, being higher redshift studies prevented by the small
statistic of the sample. The measure of the escape fraction from
typically star-forming galaxies during the
epoch of reionization will provide the missing piece of the puzzle in
our understanding of how galaxies reionize the IGM.

\subsection{Studying high-$z$ galaxies}

Theoretical models and observational evidence suggests that
most of the UV photons that reionize the Universe are emitted by low
mass, faint galaxies missed even in the deepest observations by
HST. The detection of these objects is one of the main scientific
goal of the next generation of space telescopes. However,
little or no information about the physical properties (e.g. metal and
dust content) of these objects will be accessible even with JWST.
High-$z$ GRBs can provide
a useful, complementary tool to investigate the building up of metals
at these early stages of galaxy formation and, in particular, in those
objects that provide the bulk of the ionizing photons
\citep{Salvaterra2013}. Indeed,  GRB events are produce by the same
massive stars that emit the
FUV photons able to ionize hydrogen. Metal absorption line observations of
a large sample of good signal-to-noise GRB spectra will allow to
recover the cosmic metal enrichment history, to extend the
mass-metallicity and FMR to higher redshift \citep{Salvaterra2013,Laskar2011}, to
perform stellar population studies \citep{Grieco2014,Ma2015}, and to search for peculiar
nucleo-synthesis pattern (see next Section).

Metal absorption lines over-imposed to the
afterglow radiation have been already
detected even $z>6$ showing that high-$z$ galaxies are already
enriched to a few percent solar. In particular, the spectrum of
GRB~130606A obtained with VLT/X-shooter \citep{Hartoog2014} shows a great
variety of metal elements allowing accurate abundance and gas
kinematic studies of the inner region of a $z\sim 6$ galaxy. Even better
and richer datasets will be gathered when the 30-m
class telescope like E-ELT will become operative. These studies can be complemented by X-ray spectroscopic
observations of bright GRBs with the next generation of X-ray
instruments \citep{Campana2011}. In particular, with the Athena
satellite \citep{Nandra2013} it will become possible to directly
measure the abundance patterns in X-ray afterglows allowing, in principle, to discriminate between different nucleo-synthesis
sources \citep{Jonker2013}.

Beside metal absorption lines, the observation of H$_2$
molecular absorption and of the local dust law can provide
further details about the host enrichment. As discussed in
Section~\ref{sec:obsprop}, evidence of dust extinction in four $z>6$ GRBs
has been derived by studying the X-ray-to-optical SED. Furthermore, in
the case of GRB~140515A dust deplition is suggested by the peculiar
metal abundance ratio \citep{Hartoog2014}. These observations show
that GRBs can provide fundamental clues about the presence of dust in
high-$z$ galaxies. Moreover, they can be use to constrain the dust
formation channels at work at those early epoch. It is interesting to
note that for GRB~071025 at $z\sim 5$ \citep{Perley2010} and tentatively
for  GRB~050904 at $z=6.3$ \citep{Stratta2011}  evidences for SN
synthesized dust has been reported. 
Tracking the dust enrichment history of first galaxies is also very 
important for better understand the PopIII/PopII transition. Indeed,
it has been shown that even a little amount of dust in 
the medium will induce the formation of low mass stars \citep{Schneider2003,Schneider2012}.

\subsection{Indirect search for PopIII stars}

PopIII GRBs (if any) are expected to be extremely rare and difficult
to distinguish from other GRB populations (see Section~\ref{sec:popIII}).
An indirect, but fruitful, way to search for the elusive PopIII stars
is represented by the study of the metal and dust composition of the
ISM of distant galaxies enlightened by PopII GRBs. 
Indeed, peculiar metal abundance ratios \citep{HegerWoosley2002} and dust composition \citep{Schneider2004} are foreseen as the result of the enrichment of
first massive SN explosions. The detection of such a signature
in the optical-NIR afterglow of
a very high-$z$ GRBs will provide a strong evidence for the existence
of very massive ($>100\;\Msun$) PopIII stars. 

\citet{Wang2012} found that PopIII enriched gas will result in much larger
equivalent widths of metal absorption lines than typical PopII
enrichment. Moreover, \citet{Ma2015} showed that in principle the [C/O] and
[Si/O] alone could be 
enough to distinguish a PopIII enriched envirorment, 
although, in practice, the detection of more elements
(such as S and Fe) is needed. PopII GRBs exploding in a PopIII enrichment medium are
expected to be rare at $z=6$ ($<1$\% of the GRB population at
that redshift) but the probability to find them increases
with $z$ being  $\sim 10$\% at $z=10$ \citep{Ma2015}. They
should reside in small host galaxies with $M_\star\sim
10^{5-6}\;\Msun$ and $Z<10^{-2.8}\;\Zsun$. Thus, the
detection of a $z>10$ burst with a metallicity below this threshold
will be a strong candidate to be either a PopIII event or a PopII GRBs
blowing up in a PopIII enriched medium.

\subsection{Others}

For the sake of completeness \citep[see][for a more complete list]{McQuinn2009,Amati2013} I would like to briefly mention here some
other possible uses of GRBs in the study of early Universe:

{\it - constrain the dark matter particle} \citep{Mesinger2005,deSouza2013}. The redshift distribution of GRBs can set limits
on the dark matter particle mass, m$_{\rm x}$. Using a sub-sample of
$z>4$ GRBs m$_{\rm x}>1.6-1.8$ keV at
95\% confidence level is found.

{\it - non-Gaussianity} \citep{Maio2012}. Deviations from Gaussianity of
the primordial density field will translate in different rates of GRBs
$z\gg 6$. A single GRB
detection at $z>15$ will favor
non-Gaussian scenarios with a positive non-linear parameter.

{\it - radiation field} \citep{Inoue2010,Kakuwa2012}. The ultraviolet
intergalactic radiation field below the Lyman edge energy can cause
attenuation in the $\gamma$-ray spectra of GRBs. This may
be observable at high-$z$ by the Cherenkov Telescope Array (CTA).

{\it - magnetic fields} \citep{Takahashi2011}. Pair echoes from
luminous high-$z$ GRBs may be detectable with CTA providing a unique
probe of weak intergalactic magnetic fields at early stages of
structure formation.

\section{Present and future of the fields}\label{sec:future}

The detection of high-$z$ GRBs is one of the main goals driving the
design of the next GRB missions \citep{Amati2013}. As already discussed, GRBs are
very effective in selecting high-$z$ objects and sufficiently bright to be
detectable up to beginning of the star formation activity. While the forthcoming SVOM
satellite \citep{Godet2012} or the proposed LOFT mission \citep{Amati2015} will
surely increase the $z>6$ GRB sample, it is also clear that
in order to full exploit the potentiality of GRBs as a probe of the early Univese, a much larger sample of
well studied, high-$z$ GRBs should be collected. 

\begin{figure}\label{fig:flim}
\centering
 {\includegraphics[scale=0.35]{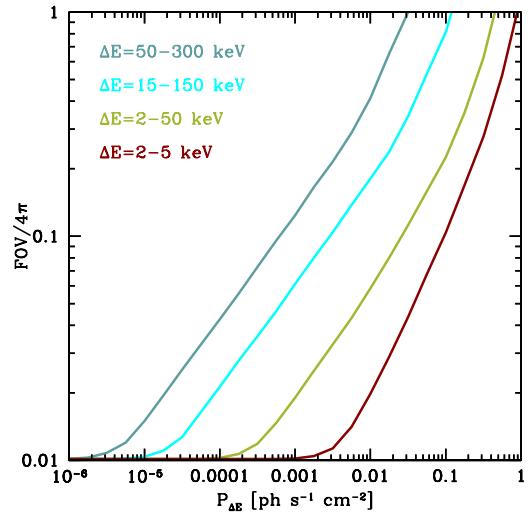}}
\caption{Required sensitivity, in terms of minimum peak flux
  $P_{\Delta E}$ that
  can be detected in the given energy band $\Delta E$, and field-of-view to
  detect 10 GRB yr$^{-1}$ at $z>8$. Different lines correspond to
  different energy bands as labeled in the plot, i.e. to different
  mission concept. See \citet{Ghirlanda2015} for the details of
the calculation.}
\end{figure}

Due to redshift scaling, in the observer frame the prompt emission of
high-$z$ GRBs is expected to peak at softer energies than lower
redshift bursts. However, as the sensitivity of current GRB satellites
allow us to detect only the bright-end of the GRB LF at
high-$z$, the mean observed peak energy is  $<E_{\rm p,
  obs}>\sim 150$ keV similar to low-$z$ bursts. Therefore, while a facility operating in the soft X-rays
will be, in general, more efficient in selecting high-$z$ GRBs, a much
better sensitivity than current instruments should also be foreseen.

The scientific goals outlined in the previous section call for a large
statistical sample of well studied high-$z$ GRBs. To be competitive
with other probes, at
least a hundred of GRBs at $z>6$ with several tens of them lying at
$z>8$ should be detected.  \citet{Ghirlanda2015} computed the expected detection rate of
high-$z$ GRBs by a generic detector with defined energy band and
sensitivity by means of a observational tested population
synthesis model of long GRBs. Following the results of this work, Figure~4
shows the sensitivity and 
field-of-view of a mission able to detect 10 GRBs yr$^{-1}$ at
$z>8$ and operating in different energy bands. For instance, by
adopting the {\it Swift} sensitivity of $\sim 0.4$ ph s$^{-1}$
cm$^{-2}$  and energy band, we can see that the goal is never reached. On the other hand, limiting ourselves to the {\it Swift} FOV
of 1.4 sr, a hundred times better sensitivity than {\it Swift} is
needed. A X-ray instrument
will instead require a sensitivity of 0.1 ph s$^{-1}$ cm$^{-2}$
in the 2-5 keV band for the same FOV. It is worth to note that the same mission will also detect
thousands of bursts at lower redshifts (plus other
transients). In order to effectively select high-$z$
candidates, the presence of 
a 0.5-1 m infrared telescope on-board should be foreseen (together
with fast repointing capabilities). This instrument will allow not
only to promptly measure the
redshift but also to perform low-resolution (R$\sim 1000$) spectroscopic studies
(e.g. to identify metal absorption lines) when the afterglow is at the
maximum of its brightness. A rapid dissemination of best high-$z$
candidates is also fundamental in order to trigger follow-up
observations by future facilities operating at different wavelengths:
optical-NIR (E-ELT, JWST if still operating), radio (SKA), X-ray
(ATHENA), TeV (CTA).

In conclusion, the study of the high-$z$ Universe with GRBs requires a
X-ray detector with unprecedented combination of
sensitivy and FOV coupled with an infrared telescope to select
 reliable high-$z$ candidates. The THESEUS mission
recently proposed for the M4 ESA call matches all these
characteristics.

\section{Conclusions}

In this paper, I have briefly reviewed the status and the prospect for
the exploration of the high redshift Universe with GRBs. 
A few solid considerations can be drawn: 

(i) GRBs do exist at very high redshift and can be detected
and studied with present-day facilities;

(ii) high-$z$ GRBs are very similar to low- and intermediate-$z$ ones;

(iii) GRBs are an efficient way to select high-$z$ objects;

(iv) high-$z$ GRBs are hosted in faint galaxies that
are missed even in the deepest surveys; 

(v) high-$z$ GRBs have proved to be an independent and powerful tool
to study the early Universe.

Although the current sample is limited to a few bursts,
future dedicated missions can provide a sufficiently large number of
GRBs at $z>6$ to study the first bound structures in a complementary
way with respect to galaxy and quasar surveys.

\section*{Acknowledgments}

This research was supported by the ASI-INAF contract (1/005/11/1). The author would like to thank A.~Melandri for providing the X-ray light
curve for the $z>6$ sample, S.~Campana for providing the measure of
the N$_{\rm H,X}$ for some of the high-$z$ bursts, G.~Ghirlanda for
provinding the $E_{\rm iso}$ values and the expected rate of GRBs at
$z>8$. 





\end{document}